\documentclass[aps,prl,twocolumn,showpacs,groupedaddress]{revtex4}  
\usepackage{graphicx}  
\usepackage{dcolumn}   
\usepackage{bm}        
\usepackage{amssymb}   

\begin{document}

\hspace{5.2in} \mbox{Fermilab-Pub-08/409-E}
\title{Search for long-lived charged massive particles with the D0 detector}

%
\author{V.M.~Abazov$^{36}$}
\author{B.~Abbott$^{75}$}
\author{M.~Abolins$^{65}$}
\author{B.S.~Acharya$^{29}$}
\author{M.~Adams$^{51}$}
\author{T.~Adams$^{49}$}
\author{E.~Aguilo$^{6}$}
\author{M.~Ahsan$^{59}$}
\author{G.D.~Alexeev$^{36}$}
\author{G.~Alkhazov$^{40}$}
% ...

\affiliation{\vspace{0.1 in}(The D\O\ Collaboration)\vspace{0.1 in}}
% ...
\affiliation{$^{6}$University of Alberta, Edmonton, Alberta, Canada,
                Simon Fraser University, Burnaby, British Columbia, Canada,
                York University, Toronto, Ontario, Canada, and
                McGill University, Montreal, Quebec, Canada}
% ...
\affiliation{$^{29}$Tata Institute of Fundamental Research, Mumbai, India}
% ...
\affiliation{$^{36}$Joint Institute for Nuclear Research, Dubna, Russia}
% ...
\affiliation{$^{40}$Petersburg Nuclear Physics Institute,
                St. Petersburg, Russia}
% ...
\affiliation{$^{49}$Florida State University, Tallahassee, Florida 32306, USA}
% ...
\affiliation{$^{51}$University of Illinois at Chicago,
                Chicago, Illinois 60607, USA}
% ...
\affiliation{$^{59}$Kansas State University, Manhattan, Kansas 66506, USA}
% ...
\affiliation{$^{65}$Michigan State University,
                East Lansing, Michigan 48824, USA}
% ...
\affiliation{$^{75}$University of Oklahoma, Norman, Oklahoma 73019, USA}
% ...
\date{September 25, 2008}

\begin{abstract}
We search for long-lived charged massive particles using $1.1$\ fb$^{-1}$ of data collected by the D0 detector at the Fermilab Tevatron $p\bar{p}$ Collider. 
Time-of-flight information is used to search for pair produced long-lived tau sleptons, gaugino-like charginos, and higgsino-like charginos. 
We find no evidence of a signal and set 95\% C.L. cross section upper limits for staus, which vary from 0.31~pb to 0.04~pb for stau masses between 60~GeV and 300~GeV. 
We also set lower mass limits of 206~GeV (171 GeV) for pair produced charged gauginos (higgsinos).
\end{abstract}
\maketitle
Charged massive stable particles, or CMSPs, are predicted by several extensions of the standard model (SM). 
The term ``stable'' in this context refers to particles that live long enough to travel several meters long and escape a typical collider detector before decaying. 
The lightest tau slepton, or stau, is an example of such a particle, and is predicted in some Gauge Mediated Supersymmetry Breaking (GMSB) models~\cite{GMSB}. 
If the stau decay is sufficiently suppressed, then the stau will be a CMSP candidate. 
The lightest chargino is another example of a CMSP. Anomaly Mediated Supersymmetry Breaking (AMSB) models~\cite{AMSB1} or supersymmetric models that do not have gaugino mass unification can predict a long lifetime for the lightest chargino if its mass is within about 150~MeV of the lightest neutralino mass~\cite{AMSB2}. 
We explore two extreme cases, one where the chargino is mostly higgsino and one where it is mostly gaugino. 

Several collider experiments have performed searches for CMSPs. 
Studies at the CERN $e^+e^-$ Collider (LEP) have resulted in lower mass limits of 97.5~GeV for stable sleptons~\cite{LEP-stau}, and 102.5~GeV for stable charginos~\cite{LEP-charginos}. 
A CDF Tevatron Run I search set a cross section limit of ${\cal{O}}(1)$ pb for stable sleptons~\cite{CDF-run1}.
Complementary searches for neutral weakly interacting massive particles (WIMPs) have also been performed by underground dark matter experiments~\cite{darkmatter}.

In this Letter, we present a search for CMSPs produced directly in pairs. We do not consider CMSPs that result from cascade decays of heavier particles. 
The detector signature of pair produced CMSPs is rather striking. 
These weakly interacting particles are expected to traverse our entire detector, and should register in its outermost muon system. 
Additionally, owing to their large mass, these particles will travel substantially slower than beam produced muons, which travel near the speed of light.

Data used in this analysis were collected with the D0 detector~\cite{run2det} at the Fermilab Tevatron $p\bar{p}$ Collider at $\sqrt{s} = 1.96$~TeV between 2002 and 2006. 
They correspond to 1.1~fb$^{-1}$ of integrated luminosity.

The D0 detector is a multi-purpose detector well suited to a wide range of searches for new phenomena. 
The main components of the detector are an inner tracker, a liquid argon and uranium calorimeter, and a muon system.
The inner tracker consists of a silicon microstrip detector (SMT) close to the beam line surrounded by a scintillating fiber detector.
The muon system~\cite{run2muon} resides beyond the calorimetry and consists of a layer of tracking detectors and scintillation trigger counters in front of 1.8~T iron toroids, followed by two similar layers after the toroids. 
Muon reconstruction at pseudorapidities~\cite{eta} $|\eta|<1$ relies on 10~cm wide drift tubes, while 1~cm mini-drift tubes are used at $1<|\eta|<2$. 
Each scintillation counter registers a passing muon's time, which can be used to calculate its speed~\cite{muongate}.
The counters have a 2--4 ns time resolution, and their calibration is maintained run-to-run within 1 ns, relative to the event time determined from the accelerator clock. 

The D0 detector uses a three-level trigger system to select data for offline analysis. 
CMSPs would appear as muons to the trigger system, so di-muon triggers were used to collect data for this analysis; and we use the term ``muon'' to refer to both real muons and CMSPs. 
Indeed, CMSPs are not distinguished from muons throughout the standard data collection and reconstruction, unless by virtue of their slow speed they arrive outside a muon trigger timing gate. The efficiency of the trigger gates is included in the calculated signal acceptance. 

Muon candidates are reconstructed by finding tracks pointing to hit patterns in the muon system. 
We select events with exactly two muons, each of which satisfies quality criteria based on scintillator and drift tube information from the muon system and matches a track in the inner tracker. 
The muon candidates are also required to have transverse momenta, $p_{T}$, greater than 20~GeV, as measured with the central tracker. 
Events with muons from meson decays and other non-isolated muons are rejected by applying the following isolation criteria. 
At least one muon must have the sum of the $p_{T}$ of all other tracks in a cone of radius ${{\cal{R}}=\sqrt{(\Delta\phi)^{2}+(\Delta\eta)^{2}}}<0.5$ around the muon direction less than 2.5~GeV. 
A similar isolation condition is applied for the total transverse energy measured in the calorimeter cells in a hollow cone of radius $0.1<{\cal{R}}<0.4$ around the muon direction; this energy must be less than 2.5~GeV. 

A cosmic ray muon that passes through the detector can be reconstructed as two collinear muons. 
To reject these events, we require that the two muons must satisfy the pseudo-acolinearity requirement
$\Delta\alpha_{\mu\mu}=|\Delta\phi_{\mu\mu}+\Delta\theta_{\mu\mu}-2\pi|>0.05$. 
Moreover, since cosmic rays can arrive at times not correlated with the beam crossing, they can be mis-identified as slow-moving particles. 
We also employ timing cuts which distinguish between outward going muons and inward traveling cosmic rays.

Two additional criteria are applied to reduce  the background from muon candidates that do not originate from the primary vertex, such as those from cosmic rays, $b$ decays, and beam halo. 
The distance of closest approach to the beam line (DCA), as measured in the transverse plane, for the track matched to the muon must be less than 0.02~cm for tracks with hits in the SMT and less than 0.2~cm for tracks without SMT hits. 
Finally, the difference in the z coordinates of the two muons at their DCAs is required to be less than 3~cm. 

We determine the total signal acceptance using a combination of information from Monte Carlo (MC) simulation and the data. 
Signal samples for CMSP masses ranging between 60 and 300 GeV were generated with {\sc pythia}~\cite{Pythia} using CTEQ6.1L~\cite{CTEQ6L1} parton distribution functions (PDF), and processed with a {\sc geant}~\cite{geant}-based simulation of the D0 detector.  
These samples were reconstructed with the same software as the data. 
The specific model used for the stable stau is model line~D in Ref.~\cite{NLSPstau}.
For long-lived charginos, the specific models follow those described in Ref.~\cite{Delphi}.
In the present analysis only direct pair production is considered so that the exact values of the model parameters of the entire supersymmetric particle mass spectrum, relevant for cascade decays, are not important.

For each scintillator layer in which the reconstructed muon has a hit, the speed of this muon is calculated and expressed in units of the speed of light. 
The average speed $\bar{v}$ is then obtained by taking the weighted average of the individual layer speeds. 
To ensure that the registered times in the muon detector are consistent, we compute a speed $\chi^{2}$ from the individual layer speeds and their uncertainties. 
We require this ${\chi^{2}}/{\it {d.o.f.}}$~to be less than 4.7, a value derived from $Z\rightarrow \mu^+ \mu^-$ data. 
The transverse momenta of the pair produced CMSPs are expected to be approximately equal, and higher than those of beam produced muons. 
To reject tracks with poorly measured momentum, we require that the absolute value of the difference over the sum of the $p_T$ of the two muon candidates in the event be less than 0.68, a value that is also derived using $Z \rightarrow \mu^+ \mu^-$ data.

Speed significance, defined as ${\left(1-\bar{v}\right)}/{\sigma_{\bar{v}}}$,
is used to distinguish slow-moving particles from near light-speed muons. 
Here $\sigma_{\bar{v}}$~is the uncertainty in the average speed $\bar{v}$. 
We require that both candidate particles in the event should have positive speed significance.

In addition to time-of-flight, we use the invariant mass formed from the pair of muon candidates to separate signal events from background. 
We calculate the invariant mass assuming the mass of each particle is that of a muon.

\begin{figure}
\begin{tabular}{cc}
\includegraphics[width=0.67\columnwidth,keepaspectratio]{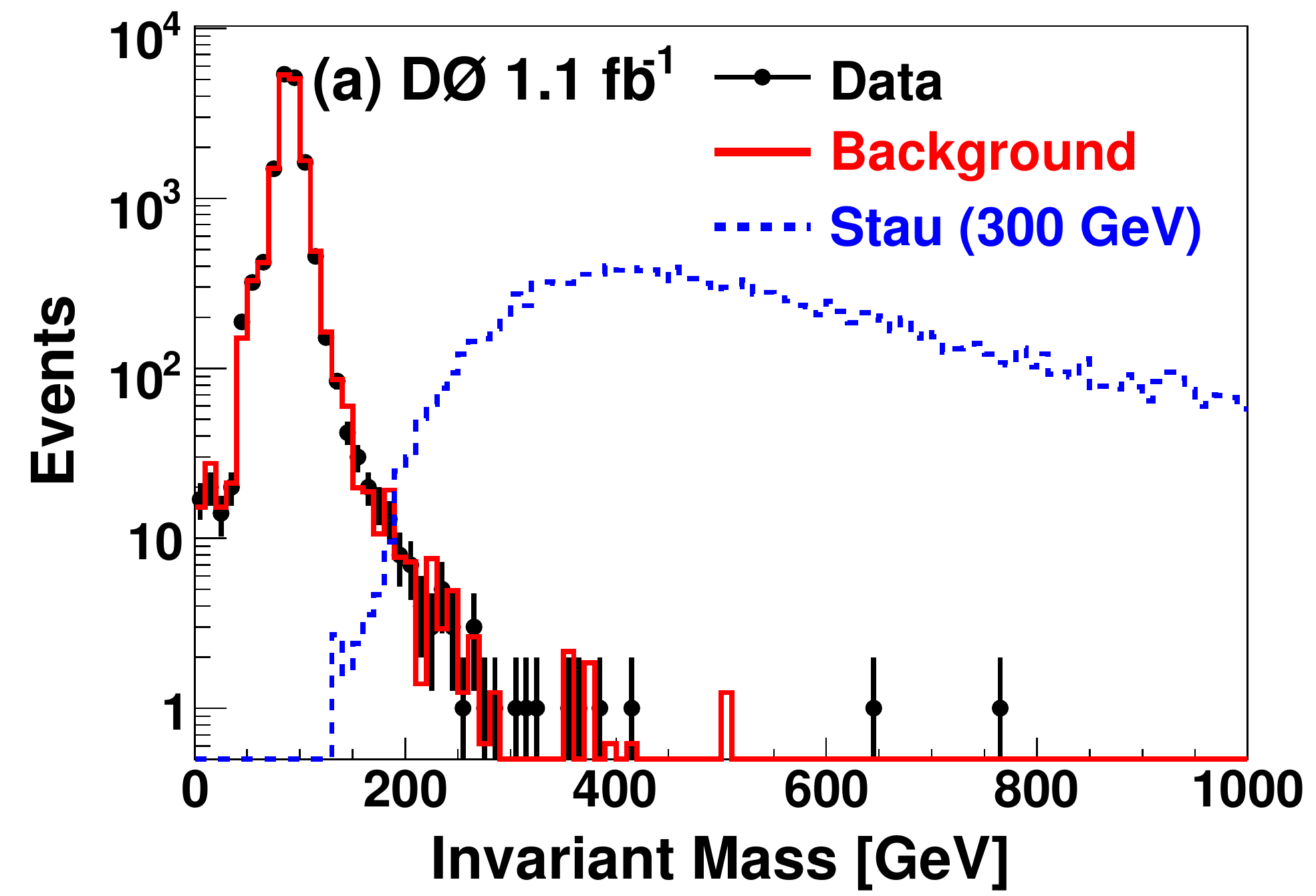}\\
\includegraphics[width=0.67\columnwidth,keepaspectratio]{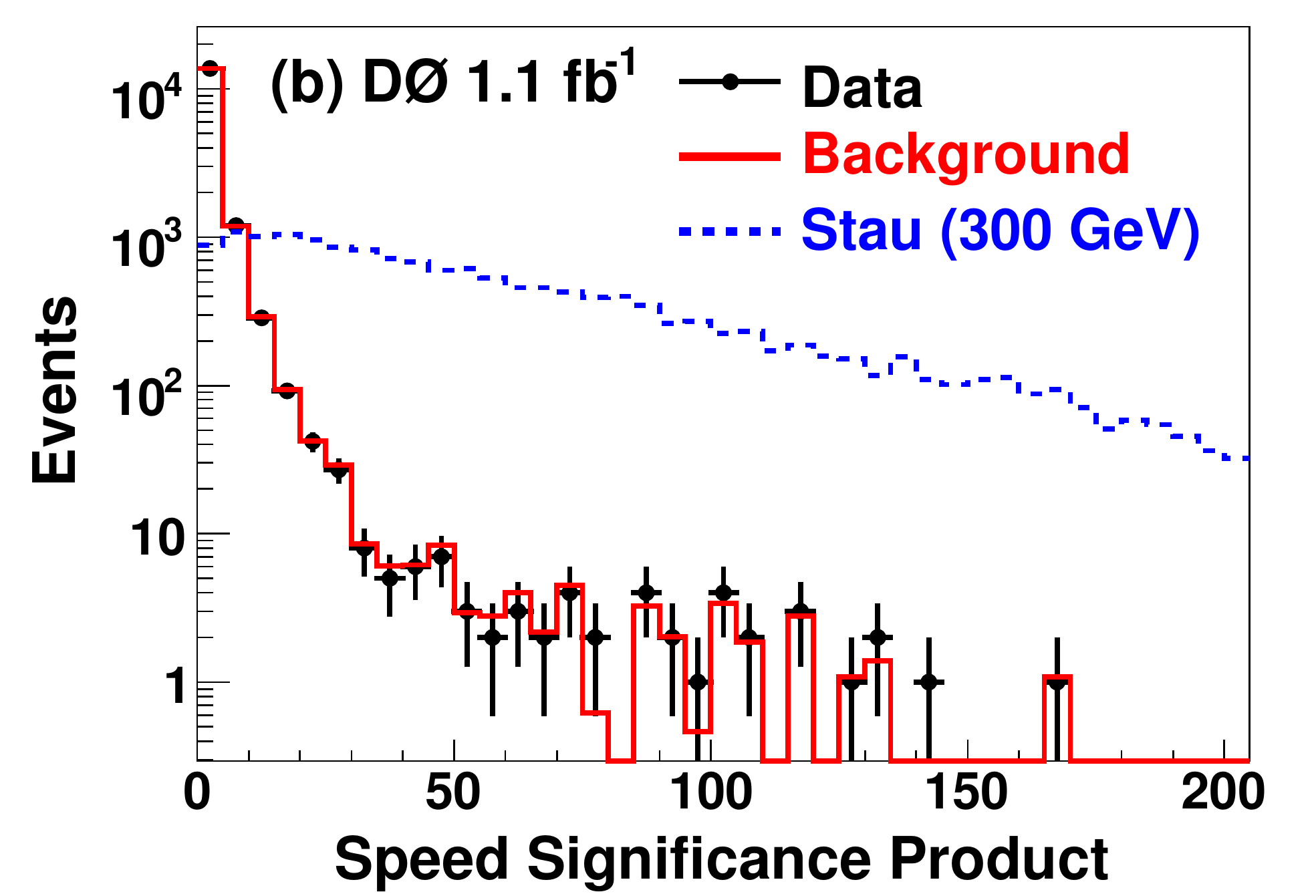}
\tabularnewline
\end{tabular}

\caption{\label{fig:lhood-input} Distributions of (a) the invariant mass and (b) speed significance product, for the simulated background (solid line), the 300 GeV stau signal (dotted line), and the data (as dots) passing the selection criteria.}
\end{figure}

The only SM background for this search comes from events which have, due to imperfect detector performance, anomalously large time-of-flight or mis-measured $p_{T}$ that satisfy the selection criteria. 
Each of these measurements is independent of the other, since the $p_{T}$ of the particle is measured in the central tracking system and the time-of-flight is recorded in the muon detector. 
Consequently, background events can be simulated by combining separate distributions of the invariant mass and of the speed significance product (the product of the values of speed significance of the two muon candidates). 
Events which pass all the selection criteria and have an invariant mass within the Z mass peak region (between 70 GeV and 110 GeV) are used to model the speed significance product distribution for the background. 
The invariant mass distribution for the background is estimated from data events that have muon candidates with negative speed significance but pass all the other selection criteria.
There is no overlap of events between these two data sets.
Background events are then simulated by choosing a random value from each of the above two distributions, the invariant mass and the speed significance product, and are normalized to the number of Z and Drell-Yan events passing the selection criteria. 
The background and MC signal samples have very different distributions, as indicated in Fig.~\ref{fig:lhood-input}. 
They are combined using a joint likelihood to discriminate between background and signal. 
The likelihood discriminant cut values are chosen for each point by minimizing the expected 95\%~C.L. upper limit on the cross section calculated with a Bayesian limit method assuming a flat prior~\cite{limit-code}. 

By construction, there is a correlation between data and the data-based simulated background in the speed significance distribution, but not in invariant mass. However, by randomly selecting from both templates, drawn from non-overlapping data sets, the variables are decorrelated for background while signal would show a correlation and peak more strongly at high likelihoods.

\begin{table}
\begin{tabular}{ccccc}
\hline \hline
\begin{tabular}{c}
Mass\\(GeV)
\end{tabular} &
\begin{tabular}{c}
Signal\\Acceptance\\(x$10^{-3}$)
\end{tabular} &
\begin{tabular}{c}
Exp.\\Signal\\Events
\end{tabular} &
\begin{tabular}{c}
Predicted\\Background
\end{tabular} &
\begin{tabular}{c}
Obs.\\Events
\end{tabular} 
\tabularnewline 
\hline
\multicolumn{4}{c} {(a) stau} \\
60&
$64\pm1\pm5$&
4.7&
$30.9\pm2.2\pm1.9$&
38\tabularnewline
80&
$38\pm1\pm5$&
1.1&
$2.6\pm0.6\pm0.4$&
1\tabularnewline
100&
$56\pm1\pm4$&
0.7&
$1.6\pm0.5\pm0.3$&
1\tabularnewline
150&
$123\pm2\pm13$&
0.3&
$1.7\pm0.5\pm0.2$&
1\tabularnewline
200&
$139\pm2\pm11$&
0.1&
$1.7\pm0.5\pm0.5$&
1\tabularnewline
250&
$133\pm2\pm13$&
0.01&
$1.7\pm0.5\pm0.3$&
1\tabularnewline
300&
$117\pm2\pm13$&
0.004&
$1.9\pm0.5\pm0.2$&
2\tabularnewline
\hline
\multicolumn{4}{c} {(b) gaugino-like charginos}\\ 
60&
$32\pm1\pm3$&
445&
$23.6\pm1.9\pm1.4$&
24\tabularnewline
80&
$24\pm1\pm3$&
85&
$1.9\pm0.5\pm0.3$&
1\tabularnewline
100&
$46\pm1\pm4$&
65&
$1.6\pm0.5\pm0.3$&
1\tabularnewline
150&
$85\pm1\pm9$&
20&
$1.2\pm0.4\pm0.1$&
1\tabularnewline
200&
$89\pm1\pm7$&
5&
$1.9\pm0.5\pm0.0$&
1\tabularnewline
250&
$74\pm1\pm7$&
1&
$1.7\pm0.5\pm0.3$&
1\tabularnewline
300&
$59\pm1\pm7$&
0.2&
$1.7\pm0.5\pm0.1$&
2\tabularnewline
\hline
\multicolumn{4}{c} {(c) higgsino-like charginos}\\ 
60&
$29\pm1\pm2$&
94&
$17.9\pm1.7\pm1.1$&
21\tabularnewline
80&
$24\pm1\pm3$&
23&
$1.6\pm0.5\pm0.3$&
1\tabularnewline
100&
$49\pm1\pm4$&
20&
$1.6\pm0.5\pm0.3$&
1\tabularnewline
150&
$89\pm1\pm9$&
7&
$1.4\pm0.5\pm0.1$&
1\tabularnewline
200&
$96\pm1\pm8$&
2&
$1.9\pm0.5\pm0.0$&
1\tabularnewline
250&
$81\pm1\pm8$&
0.5&
$1.7\pm0.5\pm0.3$&
1\tabularnewline
300&
$64\pm1\pm7$&
0.1&
$1.7\pm0.5\pm0.1$&
1\tabularnewline
\hline \hline
\end{tabular}

\caption{\label{tab:sum-cmsp}Signal acceptance, expected number of signal events, predicted number of background events and number of observed events for (a) staus, (b) gaugino-like charginos and (c) higgsino-like charginos searches, as a function of the CMSP mass. The first uncertainty is statistical and the second is systematic.}
\end{table}

The signal acceptance, expected number of signal events, predicted number of background events, and the number of observed
events are summarized in Table~\ref{tab:sum-cmsp} for staus and charginos. 
The three models studied have different signal acceptances, reflecting the different CMSP kinematics of each model.
The number of the observed events is consistent with the predicted background. 
A 95\%~C.L. upper limit on the pair production cross section is set for each mass point for the three models.

The systematic uncertainties in the background estimation arise mainly from the choice of the data events, whose invariant mass and speed significance product distributions are used to simulate the background. 
We varied the criteria used to select the data events, and the resulting difference in the predicted number of background events is taken as the size of the systematic uncertainty. 
The main signal acceptance uncertainties are those in object identification efficiencies, trigger efficiencies, MC simulation normalizations, and uncertainties related to the choice of PDF.  

\begin{figure*}[ht]
\begin{center}
\setlength{\unitlength}{1.0cm}
\begin{picture}(18.0,5.0)
\put(0.1,0.2){\includegraphics[scale=0.3]{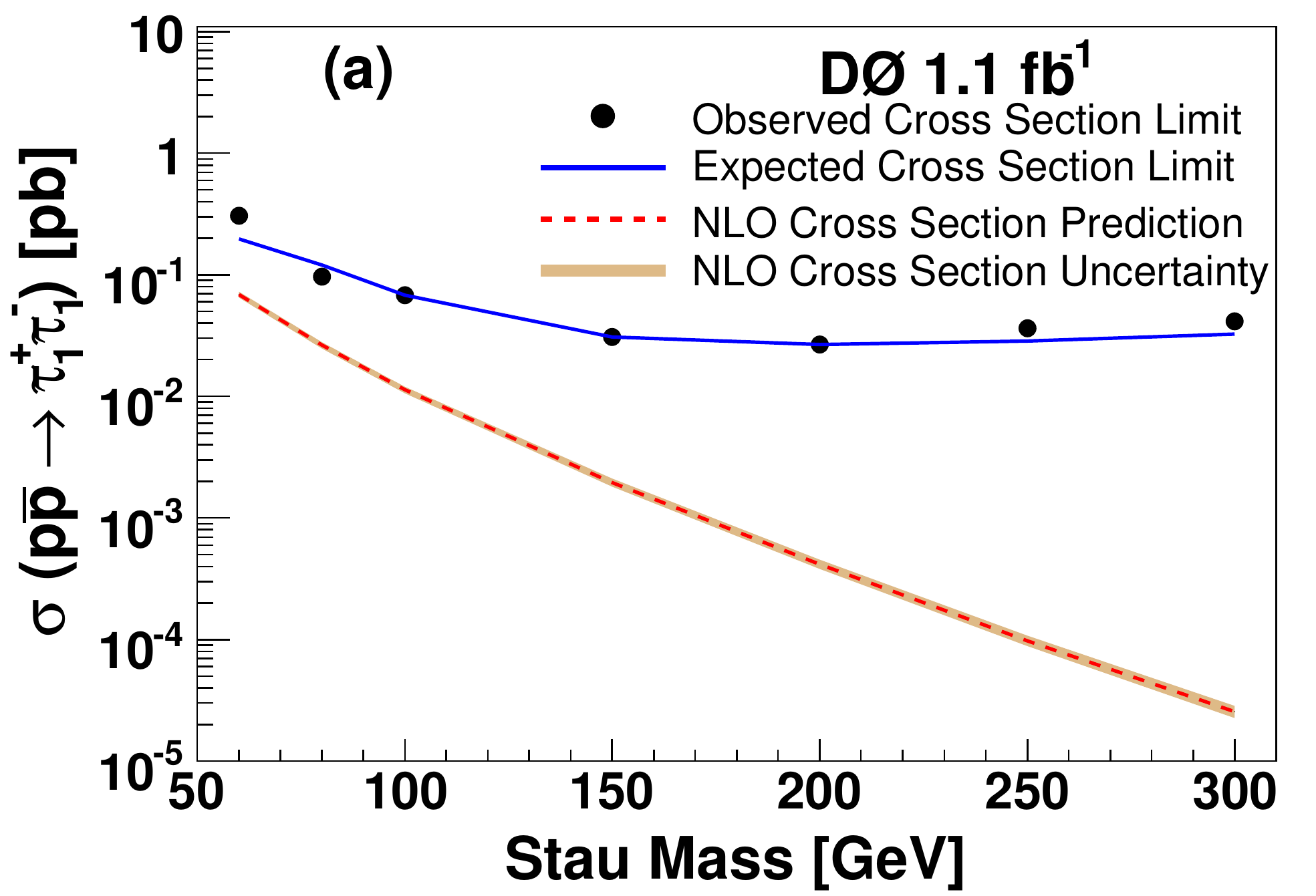}}
\put(6.1,0.2){\includegraphics[scale=0.3]{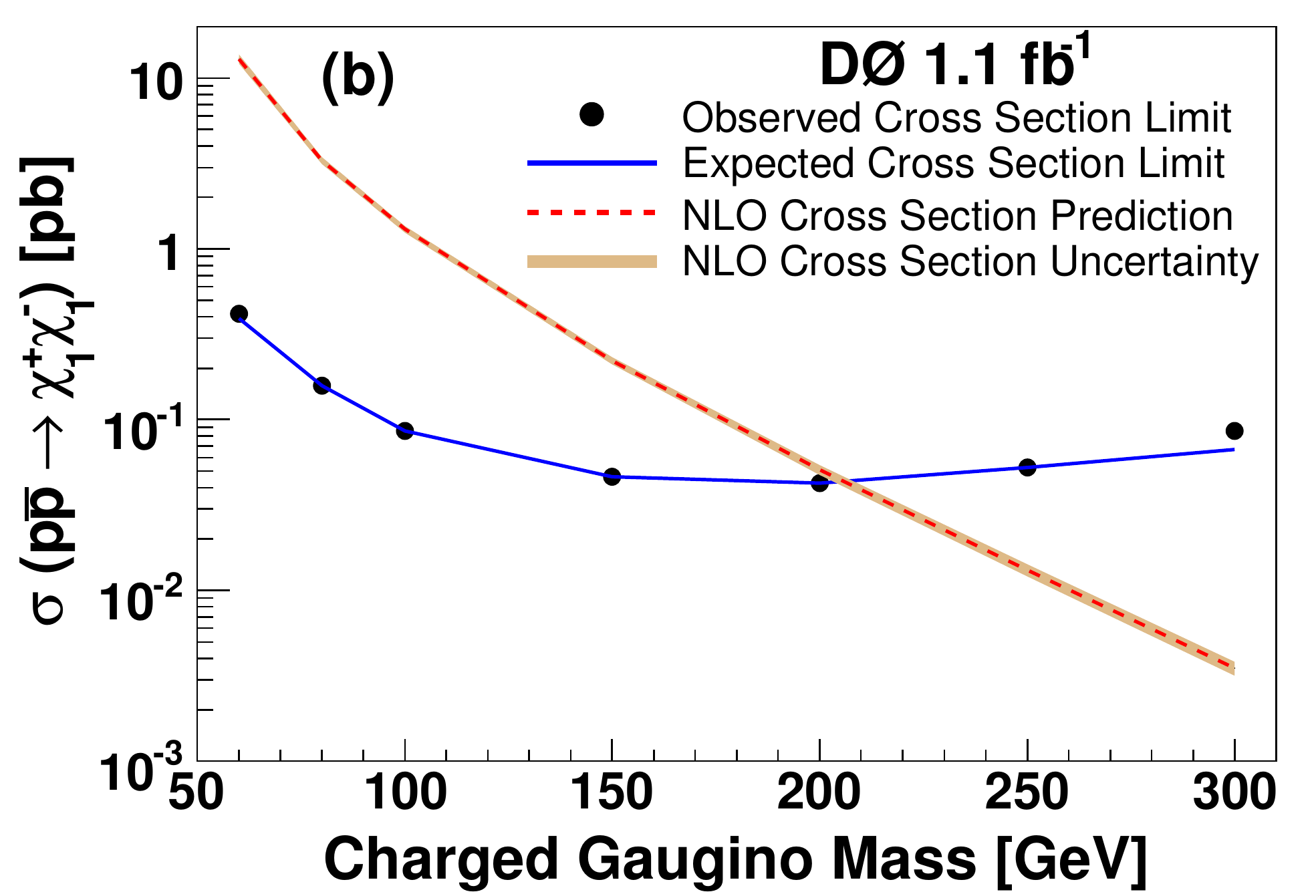}}
\put(12.1,0.2){\includegraphics[scale=0.3]{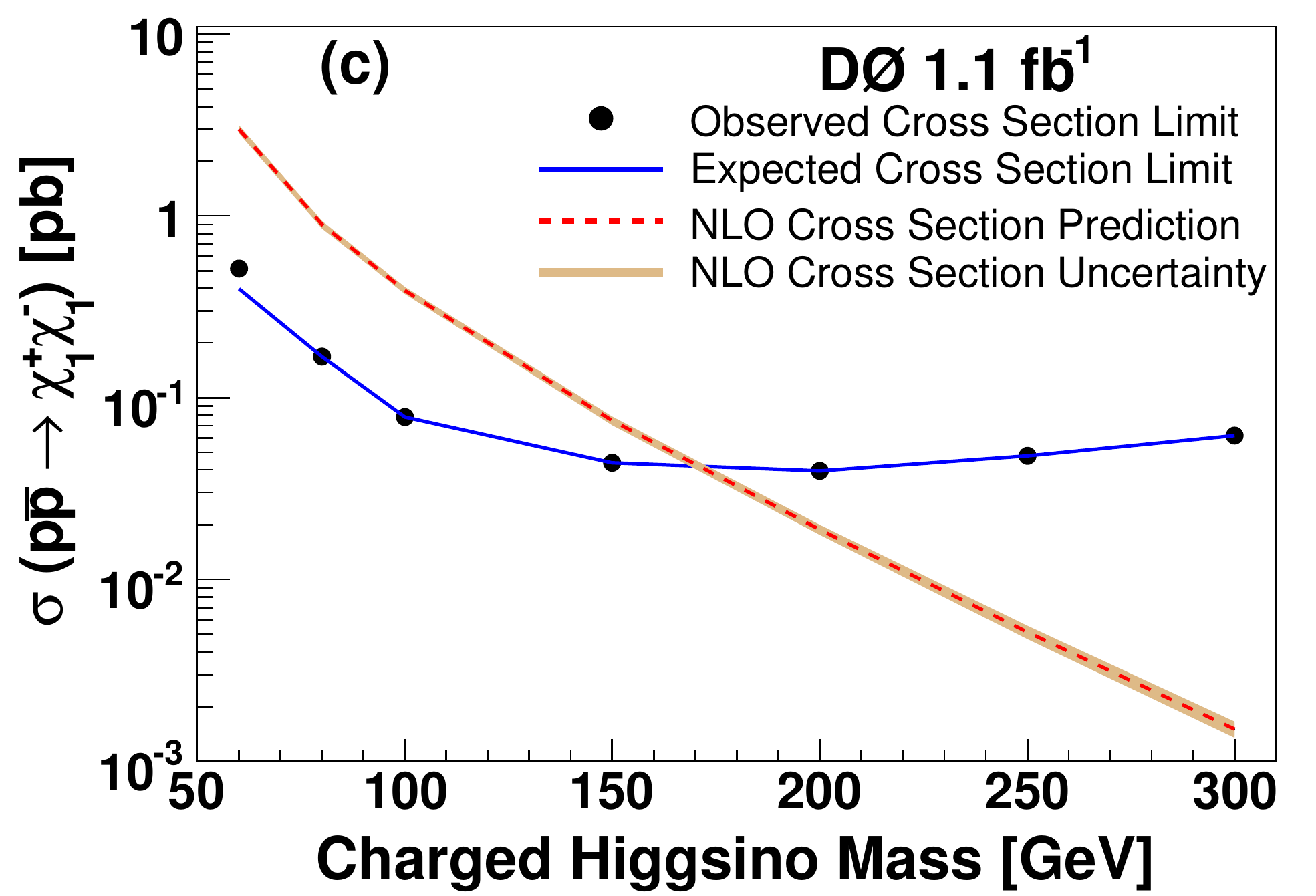}}
\end{picture}
\caption{\label{fig:limit}The observed (dots) and expected (solid line) 95\%~cross section limits, the NLO production cross
section (dashed line), and NLO cross section uncertainty (barely visible shaded band) as a function of (a) stau mass for stau pair production, (b) chargino mass for pair produced gaugino-like charginos, and (c) chargino mass for pair produced higgsino-like charginos.  }
\end{center}
\end{figure*}

The masses and couplings are computed by {\sc softsusy}~\cite{SoftSUSY}, and the next-to-leading order (NLO) cross section is calculated with {\sc prospino2.0}~\cite{Prospino}. 
The renormalization and factorization scale uncertainty and the PDF uncertainty are added in quadrature to obtain the total uncertainty on the signal cross section. 
The calculated expected and observed limits, the NLO cross section and the uncertainty on the cross section are shown in Fig.~\ref{fig:limit} for varying stau and chargino masses. 
Using the nominal (${\rm nominal}-1\sigma$) values of the NLO cross section, lower mass limits of 206 (204)~GeV at 95\%~C.L. are set for gaugino-like charginos. 
For higgsino-like charginos the limits are 171 (169)~GeV. 
Although the present sensitivity is insufficient to test the model of pair produced staus, the cross section limits can be applied to the pair production of any CMSP candidate with similar kinematics.

In summary, we have performed a search for charged massive stable particles using 1.1~fb$^{-1}$\ of data collected by the D0 detector. 
We find no evidence of a signal and set 95\%~C.L. cross section limits on the pair production of stable staus and gaugino-like and higgsino-like charginos. 
The upper cross section limits vary from 0.31~pb to 0.04~pb for stau masses in the range 60--300~GeV. 
We use the nominal value of the theoretical cross section to set limits on the mass of pair produced charginos. 
We exclude stable gaugino-like charginos with masses below 206~GeV and higgsino-like charginos below 171~GeV. These are the most restrictive limits to date on the cross sections for CMSPs and the first published from the Tevatron Collider Run II.

%
We thank the staffs at Fermilab and collaborating institutions, 
and acknowledge support from the 
DOE and NSF (USA);
CEA and CNRS/IN2P3 (France);
FASI, Rosatom and RFBR (Russia);
CNPq, FAPERJ, FAPESP and FUNDUNESP (Brazil);
DAE and DST (India);
Colciencias (Colombia);
CONACyT (Mexico);
KRF and KOSEF (Korea);
CONICET and UBACyT (Argentina);
FOM (The Netherlands);
STFC (United Kingdom);
MSMT and GACR (Czech Republic);
CRC Program, CFI, NSERC and WestGrid Project (Canada);
BMBF and DFG (Germany);
SFI (Ireland);
The Swedish Research Council (Sweden);
CAS and CNSF (China);
and the
Alexander von Humboldt Foundation (Germany).
%



\begin{thebibliography}{99}
\bibitem{eta}The D0 coordinate system is cylindrical with the $z-$axis along the proton beam direction and the polar and azimuthal angles denoted as $\theta$ and $\phi$ respectively. The pseudorapidity is defined as $\eta = - \ln[\tan(\theta/2)]$.
\bibitem{muongate} There are up to three layers of scintillation counters with each layer forming approximately the sides of a cube. The closest distance from the interaction point to each layer is 2.97 m, 4.45 m, and 5.7 m. The width of the timing gate ranges from 24 ns to 84 ns, depending on the layer.
\end{thebibliography}
\end{document}